\documentclass[a4paper]{article}
\usepackage{feynmp}
\usepackage{epsf}
\usepackage{amsmath}
\usepackage{amssymb}
\newcommand{\bloop}[3]{\fmfsurroundn{b}{#3}
\fmfpolyn{plain,smooth}{c}{#3}
\begin{fmffor}{n}{1}{1}{#3}
\fmf{phantom,tension=0.8}{c[n],b[n]}
\end{fmffor}
\begin{fmffor}{n}{#1}{#2}{#3}
\fmf{photon,tension=0}{c[n],b[n]}
\end{fmffor}
\fmffreeze
}

\renewcommand{\Re}{\mathop{\mbox{Re}}}

\newcommand{\sgn}{\mathop{\mbox{sign}}}
\newcommand{\tr}{\mathop{\mbox{tr}}}
\newcommand{\diagrambox}[3]{\raisebox{#1}{\begin{fmfgraph*}(#2)#3\end{fmfgraph*}}}
\newcommand{\myloop}[3]{\diagrambox{-4\unitlength}{10,10}{%
\bloop{#1}{#2}{#3}}}
\newcommand{\myloopII}[2]{\quad\diagrambox{-4\unitlength}{10,10}{%
\bloop{4}{6}{12}
\fmfiv{l=\ensuremath{\scriptstyle #1},l.a=0,l.d=0.5mm}{vloc(__c2)}
\fmfiv{l=\ensuremath{\scriptstyle #2},l.a=180,l.d=0.5mm}{vloc(__c8)}
}\quad}

\newcommand{\ve}[1]{\ensuremath{\mathbf {#1}}}
\newcommand{\kp}{\ve{k}}
\newcommand{\kd}{\ve{k'}}
\begin{document}
\title{Fermi gas response to non-adiabatic switching\\ of external
  potential} \author{
  Y.~Adamov \and B.~Mouzykantskii}
\date{\textit{Department of Physics, University of Warwick, Coventry,
CV4~7AL, England}}
\maketitle
\begin{abstract}
  We compute analytically the distribution function $P(E)$ for the
  energy $E$ acquired by a Fermi gas after being subjected to an arbitrary
  time-dependent external potential (switching event).  We relate the
  distribution function to a solution of a matrix Riemann-Hilbert
  problem and present explicit formulae for the low order cumulants of
  $P(E)$. These general results are used to find the distribution
  of dissipated energy in a biased quantum point contact.
\end{abstract}


\section{Introduction}
A rapid switching of an impurity potential in a Fermi gas causes
creation of electron-hole pairs which carry away some energy (so
called ``shake-up'' process). This non-trivial many body effect was
first studied more than 30 years ago in relation to the $X$-ray
absorption in metals \cite{mahan67,Nozieres69}. In this case the
perturbing potential appear instantly (i.e. the switching time $T_s$
is of the order of the inverse conduction band width $D$) and the
distribution $P(E)$ of energy $E$ carried away by electron hole pairs
is a power law $P(E) \sim E^{-\alpha}$ in the threshold region $E \ll
\hbar/T_s \sim D$.

It was later found that the same physics is important in the absorption
of ions by metallic surfaces \cite{BrakoNewns:81,Makoshi}.  In this case the
perturbing potential $V(\ve{r},t)$ changes slowly (i.e. $T_s \gg \hbar/D$).
The distribution $P(E)$ was found theoretically for arbitrary energies
\cite{BrakoNewns:81,Makoshi} (the result is also quoted in Eq.~(\ref{e-42})
below) provided that the potential $V(\ve{r},t)$ only causes  $s$-wave
scattering.

Recent advances in fabrication technology have made it possible to subject
a Fermi gas to potentials with non-trivial time and space dependence. For
example such potentials are used in adiabatic electron
pumps (see e.g.~\cite{pumps-experiment}).
 While the distribution of the pumped charge has
received a lot of attention recently \cite{ale98,AndreevKamenev:00,Levitov-cond-mat}
the closely related distribution of dissipated energy is still unknown.
In this case the time-dependent potential $V(\ve{r},t)$  changes slowly in time
(as in  the case of ion absorption) but neither the results nor the
methods from \cite{Makoshi} are applicable since other (non $s$-wave)
scattering channels are important. Finding the distribution $P(E)$ for an
arbitrary (slow) time-dependent potential turns out to be a non-trivial
problem and some novel ideas are required to solve it.

 We study the ideal Fermi gas in the presence of a
time-dependent potential $V=V(\ve{r}, t)$ assuming that the potential is
zero in the distant past ($t\rightarrow - \infty$) and becomes time
independent in the distant future. We compute the probability $P(E)$ that
the gas acquires energy $E$ after this switching process.  The results
are expressed through the scattering matrix $S(E_F,t) \equiv S(t)$ at the
Fermi energy $E_F$ on the instantaneous value of the potential $V(\ve{r},t)$.
It is more convenient to deal with the characteristic function
\begin{equation}
\label{eq:P-chi}
\chi(\lambda) = \int e^{-i E \lambda}P(E) d E
\end{equation}
rather than with the probability distribution itself. In the rest of the
paper we focus on  the case when the scattering matrix
changes slowly on the scale of Wigner's delay time, i.e.
\begin{equation}
  \label{eq:slowness}
 S^{\dagger} \frac{\partial S}{\partial E} S^{\dagger} \frac{\partial S}{\partial t}  \ll 1.
\end{equation}
Here and everywhere below we use units with $\hbar=1$. 
We start by briefly summarising the facts which are already established
in the literature (Eqs.~(\ref{eq:ND-asymptote}~--~\ref{e-42}) below).

At low energies $E \ll 1/T_s$ the distribution function has the power law
asymptote \cite{Nozieres69,Yamada-Yosida82} 
\begin{equation}
  \label{eq:ND-asymptote}
  P(E) = C (E-\Delta E)^\alpha, \quad \alpha= \frac{\tr (
  \ln^{2}S(\infty))}{4 \pi^2} -1,
\end{equation}
where $\Delta E$ is the difference between the ground state energies of
the Fermi gas with and without potential $V(\ve{r},\infty)$~\cite{Mahan-book} and is
given by
\begin{equation}
  \label{eq:E0}
 \Delta E=\frac{i}{2\pi}\int_{-\infty}^{0}dE \tr\ln S(E,\infty).
\end{equation}

The average absorbed energy $ \langle E \rangle = \int P(E) E dE $ and
the probability $P_{00}$ to remain in the ground state are given by
\begin{eqnarray}
\label{eq:loss}
\langle E \rangle &=&-\frac{1}{4\pi}\int dt \tr\{(\partial_{t}S)
S^\dagger(\partial_{t}S) S^\dagger\}, \\
  \label{eq:P00}
  P_{00}&=&\exp\left\{\frac{1}{4\pi^2}\int_0^{\infty} \tr
    \left(D(\omega)D(-\omega)\right) \omega 
    d\omega \right\}
\end{eqnarray}
respectively \cite{Makoshi:83}, where $D(\omega)=\int e^{i \omega t} \ln S(t) dt$. 
When $V(\ve{r},\infty) \ne 0$ the integral in~(\ref{eq:P00}) diverges
logarithmically giving rise to the Anderson orthogonality catastrophe
\cite{Anderson-67}. 

Finally, in the ``commutative case'' when the scattering matrix $S(t)$ can
be diagonalised in a time-independent basis (e.g. when there is only
$s$-wave scattering) the problem admits a complete solution and the
characteristic function is given by \cite{Makoshi,BrakoNewns:81}
\begin{equation}
  \label{e-42}
    \ln\chi(\lambda) =-i \Delta E \lambda +
 \frac{1}{4\pi^2}\int_0^{\infty}
 |D(\omega)|^2 \omega (1-e^{-i\omega \lambda}) d\omega  .
\end{equation}

To illustrate the above results and compare them with our findings we
consider a simple model of a quantum point contact biased by a
rectangular voltage pulse $V_b(t) = V_0 \theta(t)
\theta(T_s-t)$. After a gauge transformation (see Sec.~\ref{s-72} for
details) the potential becomes local in space and the scattering
matrix takes the form
\begin{equation}
\label{eq:2x2S}
S(t)=\left[ 
 \begin{array}{cc}
  Ae^{i \phi(t)} & -B^{*}\\
  B & A^{*}e^{-i\phi(t)}
 \end{array}
\right],
\end{equation}
where $A,B$ are transmission and reflection amplitudes  and
 $\phi(t)=\int_{-\infty}^{t}eV_b(\tau)d\tau$ is the Faraday flux.

The results~(\ref{eq:ND-asymptote}--\ref{e-42}) are of not much use for
the simple model introduced above:  since the matrices in~(\ref{eq:2x2S}) do
not commute with each other at different times, Eq.~(\ref{e-42}) is not
applicable and the only known result is Eq.~(\ref{eq:loss}) for the
average energy.  This result, however, is just a straightforward
combination of the Joule law $ \langle E \rangle = \int I(t) V_b(t) dt
$ with the Landauer-B\"uttiker formula \cite{StandardButtiker,Landauer85} for the
current across the junction $ I = \frac{e^2}{2 \pi} |A|^2 V_b$.

In this paper we develop a technique for finding distribution function
$P(E)$ which allows us\\
-- to  obtain the cumulants expansion of $P(E)$ explicitly and to
compute the
second and third cumulants for an arbitrary switching process
(see Eqs.~(\ref{e-2nd},~\ref{e-3rd}))\\
-- to relate $P(E)$ to a solution of a matrix Riemann-Hilbert (RH)
problem (Eq.~(\ref{eq:RH}))\\
-- to find the distribution function of energy dissipation in the quantum
point contact biased by a long rectangular pulse (see below)

We finish the introduction by illustrating the relation between $P(E)$
and the statistics of transmitted charge. Consider a quantum point
contact described above. When the rectangular pulse is long i.e.  $T_s
\gg 1/(e V_0)$ we solve the RH problem asymptotically to
logarithmic accuracy in $n=eV_0 T_s/(2\pi) \gg 1$; this gives
\begin{eqnarray}
\label{eq:largeTans}
\ln\chi(\lambda) &=& n\ln(|A|^{2}e^{-i e
V_0\lambda}+|B|^{2})+\nonumber\\
 &+& \frac{1}{2\pi^{2}}
\ln(\frac{2\pi n}{e V_0 \lambda})\ln^{2}(|A|^{2}e^{-i e
V_0\lambda}+|B|^{2}),   
\end{eqnarray}
where we assume $eV_0\lambda \ll n$.  Since each electron  reaching
the contact contributes $e V_0$ to the total dissipated energy if it is
transmitted (probability $|A|^2$) and contributes nothing if it is
reflected (probability $|B|^2$) we expect at least for long times $T_s$
that characteristic function~(\ref{eq:largeTans}) can be obtained
from the distribution of transmitted charge. The latter is known to be
binomial \cite{LesovikLevitovJETP93}
\begin{equation}
\label{e-59}
B(N,k)=\frac{N!}{k! (N-k)!}
|A|^{2k}|B|^{2(N-k)}
\end{equation}
where $B(N,k)$ is the probability that out of $N$ incident electrons
$k$ are transmitted.  The distribution $p(N)$ of the number of incident
electrons can be found from the energy dissipation of a completely
open contact (i.e. one with $A=1$). The mean number of electrons 
transmitted through an open
contact is given by $\langle N \rangle = n$, and the distribution
$p(N)$ is Gaussian in the vicinity of its maximum (see Sec.~\ref{s-trcont} for
details) 
\begin{equation}
\label{e-60}
p(N)\sim e^{-\pi^2 (N-n)^2/(4 \ln n)} \quad \hbox{at} \quad n-N \sim 1.
\end{equation}
There are therefore two sources of shot noise in a quantum point
contact~-- one is the fluctuations in the number of incident electrons
and the other is the fluctuations in the number of reflected
electrons. We are now able to establish that these two sources are
statistically independent by observing that the characteristic
function~(\ref{eq:largeTans}) corresponds (with logarithmic accuracy)
to the convolution of $p(N)$ and $B(N,k)$:
\begin{equation}
P(E)=\sum_{N,k}p(N)B(N,k)\delta(E-eV_0 k).
\end{equation}
The physical picture of the two statistically independent sources of
quantum shot noise was suggested (but not proven) in \cite{Levitov94}.
Note, that in order to confirm the statistical independence we need to
obtain the sub-leading term in the large $n$ expansion of $\chi(\lambda)$
(namely the term proportional to $\ln n$ in Eq~(\ref{eq:largeTans})) and
analyse its dependence on the transmission amplitude $A$. We are not
aware of any other technique capable of obtaining this term
for a partially open channel.

In fact, measuring the energy dissipated in a contact in the presence of
a rectangular bias pulse can be viewed as a ``quantum charge measurement
scheme'' alternative to (and somewhat conceptually simpler than) using a
precessing spin as was suggested in \cite{LesovikLevitovJETP93}. The detailed
comparison of the two measurement schemes will be the subject of future
work.

\section{Overview of the method}

Let $ H(t) = H_0 + V(\ve{r},t)$ be the time dependent Hamiltonian of the
Fermi gas in the external potential, where $H_0$ is the Hamiltonian of
the free Fermi gas. 
 
If $V(\ve{r},\infty) \neq 0$ then there are two contributions to the
energy of the gas at $t\to\infty$. The first is the sum of the
energies of the electron-hole pairs and the second originates from the
adiabatic shift of the single particle energy levels.  If the energy
levels in the vicinity of the Fermi surface shift uniformly by
$V(\ve{r},\infty)$ then we can adiabatically switch off the potential
$V(\ve{r},\infty)$ preserving the energy of existing electron-hole
pairs and creating no additional ones. Therefore to account for
$V(\ve{r},\infty)$ we should add $-i\Delta E\lambda$ to the
characteristic function $\chi(\lambda)$.  We assume
$V(\ve{r},t>T_s)=0$ in the rest of the paper.

The evolution operator $ U_t $ which relates the many-body wave function
$|\Psi(t)\rangle$ to the the ground state wave function $|\Psi_0\rangle$
via $|\Psi(t)\rangle = U_t |\Psi_0\rangle$ obeys the Schr\"odinger
equation
\begin{equation}
  \label{eq:U}
  i \frac{d}{dt} U_t  = H(t)  U_t.
\end{equation}
The characteristic function~(\ref{eq:P-chi}) is given by the average
\begin{equation}
 \label{e-chi-lambda}
 \chi(\lambda)=
 \lim_{t \to \infty} \langle \Psi_0 | U_t^{-1} e^{-i H_0 \lambda}
 U_t e^{i H_0 \lambda}|\Psi_0\rangle.
\end{equation}
Usually such averages are evaluated by expanding both $U$ and $U^{-1}$ in
the powers of $V(\ve{r},t)$ and by evaluating the result using
the Keldysh diagram technique (see \cite{Makoshi} for application of
this method in the commutative case). We follow a different approach
suggested in \cite{BrakoNewns:81} and separate the single-particle
scattering problem from the averaging over the Fermi-gas ground state.
The scattering problem consists of finding the electron annihilation
operator $\hat c_\kp(t)$ in the Heisenberg representation
\begin{equation}
\label{e-44}
\hat{c}_{\kp}(t)=U_{t}^{-1}c_{\kp}U_{t} = e^{-i
  \epsilon_{\kp} t} \sum_{\kd}\sigma_{\kp\kd}(t)c_{\kd},
\end{equation}
where $\epsilon_{\kp}$ is the energy of the electron with momentum
$\kp$.  The matrix $\sigma_{\kp\kd}(t)$ becomes time independent at $t >
T_s$ when the potential is switched off. Only this limiting value is
needed for computation of the characteristic function~(\ref{e-chi-lambda}).
It can be expressed as a Fourier component of the scattering matrix
\begin{equation}
\label{e-43}
\sigma_{\kp\kd}(\infty)=
\frac{\pi^{d/2-1}}{\Gamma(\frac{d}{2})\nu(E)}\int dt S_{\ve{n}\ve{n'}}(E,t)
e^{i(\epsilon_{\kp}-\epsilon_{\kd})t}
\end{equation}
where $\nu(E)$ is the density of states, $d$ is the number of space
 dimensions, $\ve{n}=\kp/|\kp|$ and $E=(\epsilon_{\kp}+\epsilon_{\kd})/2$.
Eq.~(\ref{e-43}) is the first term in the expansion in
 $(S^{\dagger}\frac{\partial S}{ \partial E} S^{\dagger}
\frac{\partial S}{ \partial t})$ and is valid as long as ($\epsilon_\kp -
\epsilon_\kd)S^{\dagger}\frac{\partial S}{ \partial E} \ll 1$.
 The detailed calculation  is presented in appendix~\ref{s-1part}.

Using~(\ref{e-44}),(\ref{e-43}) we can compute the limiting value of the
Hamiltonian in the Heisenberg representation
\begin{equation}
\hat{H}(t_\infty)=U_{t_\infty}^{-1}H_{0}U_{t_\infty}= 
\sum_{\kp}\epsilon_{\kp}c^{\dagger}_{\kp}c_{\kp}+
\sum_{\kp,\kd}c^{\dagger}_{\kp}W_{\kp\kd}c_{\kd},
\end{equation}
where 
\begin{equation}
\label{eq:wkp}
W_{\kp\kd}=\frac{\pi^{d/2-1}}{\Gamma(\frac{d}{2})\nu(E)}
\int i S^{\dagger}(E,t)\frac{\partial S(E,t)}{\partial t}
e^{i(\epsilon_{\kp}-\epsilon_{\kd})t}dt.
\end{equation}

\begin{fmffile}{ring1}
Now the characteristic function~(\ref{e-chi-lambda}) acquires the form of
a time independent expectation value and can be evaluated using
a linked-cluster expansion (see appendix~\ref{s-68} for details) 
\begin{equation}
\label{e-54}
\ln\chi(\lambda)=\frac{i\lambda}{4\pi}\int dt \tr\{(\partial_{t}S)
S^\dagger(\partial_{t}S) S^\dagger\}+{\mathcal W}
\end{equation}
where $\mathcal W$ is the sum of the ring diagrams starting with the
second order term
\begin{equation}
\label{e-diagrams2}
{\mathcal W}=-\frac{1}{2}\myloop{4}{6}{12}-\frac{1}{3}\myloop{4}{4}{12}-\cdots
\end{equation}
Here the  Green's function is 
\begin{equation}
\label{e-greenwz}
G(p,\Delta\tau)=-ie^{i p\Delta\tau}\left(\theta(-p)\theta(\Delta\tau)-
\theta(p)\theta(-\Delta\tau)\right) 
\end{equation}
and the  vertex is given by
$A(p-p',\tau)=M(p-p')(\theta(\tau)-\theta(\tau-\lambda)$ where 
\begin{equation}
\label{e-50}
M(p-p')=i\int  S^{\dagger}(t)\frac{\partial S} {\partial t} e^{-i
(p-p') t}dt.
\end{equation}  
We integrate over $\tau$ and sum over matrix indices of $M(p-p')$
in each vertex and sum over $p$ in each line.

The diagram expansion~(\ref{e-diagrams2}) makes the computation of
the $n$-th order cumulant
\begin{equation}
\label{e-nth}
\langle\langle E^n\rangle\rangle=i^{n}\frac{\partial^{n}\ln\chi(\lambda)}{\lambda^n}
|_{\lambda=0}
\end{equation}
relatively straightforward. Indeed, since the $n$-th diagram is
proportional to $\lambda^n$ at small $\lambda$ no more than $n$ terms is
needed for the $n$-th cumulant. This gives  the following results for
the second and the third cumulants
\begin{eqnarray}
\label{e-2nd}
\langle\langle E^2 \rangle\rangle&=&\frac{1}{4 \pi^2}\int_{0}^{\infty}
  \tr \left(M(p)M(-p)\right) p dp \\
\label{e-3rd}
\langle\langle E^3 \rangle\rangle &=& \frac{1}{
  (2\pi)^3}\int_{0}^{\infty} \min(p_1,p_2)\tr\left(M(p_2-p_1)
[M(-p_2),M(p_1)]\right)dp_{1} dp_{2}+
\nonumber\\
 &+& \frac{1}{4\pi^2}\int_{0}^{\infty}
\tr\left(M(p)M(-p)\right) p^{2} dp.
\end{eqnarray}

Finding the characteristic function $\chi(\lambda)$ is a much more
difficult problem which we relate to  the
RH problem  described below.  We start by introducing the matrix
valued function $S_+(t,y)$ which arbitrary interpolates between $1$ at
$y=0$ and $S(t+\lambda)$ at $y=1$
\begin{equation}
S_{+}(t,1) = S(t+\lambda), \quad S_{+}(t,0)=1
\end{equation}
and analogously the matrix valued function $S_{-}(t,y)$
interpolating between $1$ and $S(t)$
\begin{equation}
S_{-}(t,1) = S(t), \quad S_{-}(t,0)=1.
\end{equation}
It is also convenient to define the new variables $z=\tau+it$ and
 $\bar{z}=\tau-it$.
For a fixed $0<y<1$ the RH problem consists of finding two
 matrix valued functions $f_{\pm}(\bar{z},y)$
that satisfy the following conditions:\\
(i) $f_{+}(\bar{z},y)$ is antianalytic when $\Re \bar{z} >0$\\
(ii)$f_{-}(\bar{z},y)$ is antianalytic when $\Re \bar{z}<0$\\
(iii)$f_\pm(\infty,y)=1$\\
(iv) for real $t$
\begin{equation}
\label{eq:fpm}
f_{+}(-it+0,y)f_{-}(-it-0,y)^{-1}=Q(t,y),
\end{equation}
where  $Q=S_{+}^{-1}S_{-}$.

 Once the solution of the RH problem (or approximation to it) is
found the characteristic function of dissipated energy is given by the
two-dimensional integral
\begin{eqnarray}
\label{eq:RH}
\ln\chi(\lambda) &=& \frac{i}{4\pi}\int
\tr\left\{dS_{+}S_{+}^{-1}\wedge
dS_{-}S_{-}^{-1}\right\}\nonumber\\
 &-& \frac{i}{4\pi}\int
\tr\left\{Q^{-1}dQ\wedge df_{-}f_{-}^{-1}\right\},
\end{eqnarray}
where $\wedge$ is the wedge product, the integrals are over the strip
($0<y<1, -\infty <t< +\infty$) and the orientation is chosen in such a
way that $\int dy \wedge dt$ is positive.

The difficult step in the outlined procedure is the solution of the RH
 problem. We consider two special cases. In the quantum point contact
 case with a particular $t$-dependence of $S$-matrix  the
 RH problem can be solved asymptotically(see
 section~\ref{s-45}).
 The ``commutative'' case when the $S$ matrix can be diagonalised in
some time-independent basis is discussed below.  For pure $s$-wave
scattering the $S$ matrix has only one nontrivial eigenvalue
$S(t)=e^{-2i\delta_V(t)}$, where $\delta_V(t)$ is the phase shift at
the Fermi energy. The addition of more scattering channels is
straightforward. Formulating~(\ref{eq:fpm},~\ref{eq:RH}), 
we choose the interpolating functions $S_\pm$ in the form
\begin{equation}
  \label{e-20}
S_-(t,y)=e^{-2i y\delta_V(t)}, \quad S_+=e^{-2i y\delta_V(t+\lambda)}
\end{equation}
which gives $Q=e^{-2i y(\delta_V(t)-\delta_V(t+\lambda))}$. Employing
the method from \cite{Muskhelishvili} we  obtain the solution
\begin{equation}
\label{e-22}
\ln f_{\pm}(\bar{z},y)=
\frac{i y}{\pi}\int d t_{1}\frac{\delta_{V}(t_{1})-\delta_{V}(t_{1}+\lambda)}
{\bar{z}+i t_{1}\pm 0}.
\end{equation}
Substituting~(\ref{e-22},\ref{e-20}) into~(\ref{eq:RH})
gives~(\ref{e-42}).

Equations~(\ref{e-2nd}),~(\ref{e-3rd}),~(\ref{eq:RH})
and Eq.~(\ref{eq:largeTans}) from the introduction are the main
results of the paper. 

\section{Wess-Zumino action and the Riemann-Hilbert problem}

In this section we employ the Wess-Zumino action to
find $\mathcal W$  given by the diagram 
series~(\ref{e-diagrams2}). 
Firstly, we show that the functional derivative of $\mathcal W$ with
respect to the vertex $A$
\begin{equation}
\label{e-48}
j=2i\frac{\delta {\mathcal W}}{\delta A}=-2iGAG-2iGAGAG-\cdots  
\end{equation}
satisfies the equation  for the current in the Wess-Zumino model.
Observe that the Green's function $G(p,\tau)$ given by Eq.~(\ref{e-greenwz})
obeys the same equation as the Green's function of a massless
left-moving particle in  one dimension,  $p$ being  the
momentum of the particle and $x$  is the conjugate variable to $p$  
\begin{equation}
\label{e-47}
(\partial_{\tau}-\partial_{x})G(x-x',\tau-\tau')=-i\delta(x-x')\delta(\tau-\tau').
\end{equation}
Note, that in the coordinate representation the vertex $A$ takes the
form
\begin{equation}
A(\tau,x)=iS^{\dagger}\frac{dS}{dx}\left(\theta(\tau)-\theta(\tau-\lambda)\right).
\end{equation}

Using~(\ref{e-47}) we get for the
$n$-th order $(n>2)$ of the  perturbation expansion~(\ref{e-48})
\begin{equation}
\label{e-51}
(\partial_{\tau}-\partial_{x})j_{n}=-i[A,j_{n-1}].
\end{equation} 
 The second order diagram can be presented in the form
$$
-\frac{1}{2}\int\frac{dp}{2\pi}\frac{d\omega}{2\pi}
\Pi(p,\omega)A(p,\omega)A(-p,-\omega)
$$
where $\Pi(p,\omega)=\frac{-ip}{2\pi(\omega+p-i0\sgn p)}$ 
is a polarisation operator. Varying this equation over $A$ we have the
following equation for $j_2$
\begin{equation}
\label{e-73}
(\partial_{\tau}-\partial_{x})j_{2}=
\frac{-1}{\pi}\partial_{x}A.
\end{equation}

Combining~(\ref{e-51}) and~(\ref{e-73}) we arrive at the  implicit
equation for $\mathcal W$
\begin{equation}
\label{e-fakeWZ}
(\partial_{\tau}-\partial_{x})j+i[A,j]=
\frac{-1}{\pi}\partial_{x}A; 
\quad j=2i\frac{\delta \mathcal W}{\delta A}.
\end{equation}
 This equation is very similar to the equation for currents in
the Wess-Zumino model. To exploit this analogy  we introduce the functional 
\begin{equation}
\label{e-74}
\tilde{\mathcal W}={\mathcal W}+\frac{i\lambda}{8\pi}\int
\tr(\partial_x S S^{-1})^2 dx={\mathcal W}-\frac{i}{8\pi}\int \tr
A^2(x,\tau)dx d\tau
\end{equation}
which obeys the implicit equation
\begin{equation}
\label{e-8}
(\partial_{\tau}-\partial_{x})\tilde{j}+i[A,\tilde{j}]=
\frac{-1}{2\pi}(\partial_\tau+\partial_{x})A; 
\quad \tilde{j}=2i\frac{\delta \tilde{\mathcal W}}{\delta A}.
\end{equation} 
 
Polyakov and Wiegman(PW) \cite{PolyakovWiegman:84} solved~(\ref{e-8})
for the case when $A$ has the form
\begin{equation}
\label{e-53}  
A=i(\partial_{\tau}-\partial_x)g g^{-1}; 
\quad g|_{\tau\rightarrow\infty, x\rightarrow\infty}=1,
\end{equation}
 where $g$ takes values
in  $SU(N)$.  
Under these conditions Eq.~(\ref{e-8}) has the solution 
\begin{eqnarray}
\label{e-55}
\tilde{\mathcal W}(A) &=&
-\frac{i}{8\pi}\int dx d\tau\tr\{\partial_{\mu}g g^{-1}
\partial^{\mu}g g^{-1}\}\nonumber\\
 &-& \frac{i}{12\pi}\int dx d\tau dy \epsilon^{\mu\nu\lambda}
\tr\{\partial_{\mu}g g^{-1} \partial_{\nu}g g^{-1}
\partial_{\lambda}g g^{-1}\}. 
\end{eqnarray}
In the second term in~(\ref{e-55}) the integration is over a membrane
in the target space with boundary $g(\tau,x)$. The current
is given by $\tilde{j}=\frac{-i}{2\pi}(\partial_{\tau}+\partial_{x})g g^{-1}$.

In our case the PW procedure  is not
applicable because the vertex $A$ does not admit 
representation~(\ref{e-53}). This can be demonstrated by using
Eq.~(\ref{e-48}) for the current and  observing that  it does not
decay  at $x,\tau \to \infty$, in contradiction to the PW prediction. For
example, the second term in~(\ref{e-48}) gives  
$j_2=2i\int\frac{1}{|x-x'+\tau-\tau'-i0sgn(\tau-\tau')|^2}A(\tau',x')dx'd\tau'$,
which is not zero if $x \to\infty$, $\tau\to\infty$ along the line
$x=-\tau$.

Boundary conditions that are more general than~(\ref{e-53})  were
considered in \cite{Falomir90,Falomir91} using a rather involved method.  
Instead, we overcome the problem by rotating the contour of integration over
the energy $\epsilon$ in  diagram expansion~(\ref{e-diagrams2})(the
procedure is  clearer in the energy representation) and        
take  $\epsilon=i\xi$ where $\xi$ is real and varies from
 $-\infty$ to $\infty$. 
As the Fourier transform of $[\theta(\tau)-\theta(\tau-\lambda)]$ is
$\frac{e^{i(\epsilon-\epsilon')\lambda}-1}{i(\epsilon-\epsilon')}$,
 we also need to rotate $\lambda$ to keep the integrals
convergent. In this way we obtain $\chi(\lambda)$ at
$\lambda=-i\alpha$; where $\alpha$ is real and positive. The
characteristic function for a real $\lambda$ is obtained using
analyticity of $\chi(\lambda)$ in the lower half-plane.  
The vertex in the coordinate representation \cite{coords} is
$A(\tau,x)=iS^{\dagger}(x)\frac{\partial S(x)}{\partial x}
(\theta(\tau+\alpha)-\theta(\tau))$ and the rotated Green's function is
$G(x-x',\tau-\tau')=\frac{1}{2\pi(\tau-\tau'-i(x-x'))}$. This Green's
function satisfies equation
$2\partial_{z}G=\delta(x-x')\delta(\tau-\tau')$ where we denote
$z=\tau+ix$.

 Similar arguments as above lead to the equation
for the current  $\tilde{j}=2i \frac{\delta\tilde{\mathcal W}}{\delta A}$ (note that the
current now goes to zero as $z\to\infty$)
\begin{equation}
\label{e-49}
2\partial_{z}\tilde{j}-[A,\tilde{j}]=\frac{-i}{\pi}\partial_{\bar{z}}A.
\end{equation} 
Equation~(\ref{e-53}) now takes the form 
\begin{equation}
\label{e-9}
2\partial_{z}g g^{-1}=A=iS^{\dagger}(x)
\frac{\partial S(x)}{\partial x}
(\theta(\tau+\alpha)-\theta(\tau))
\end{equation} 
and has the solution with $g\in GL(N,{\mathbb C})$
 satisfying $g|_{z\rightarrow\infty}=1$,
\begin{equation} 
\label{e-12}
g(\tau,x)=\left\{
\begin{array}{ll}
f(\bar{z}), & \tau<-\alpha,\tau>0\\
S^{\dagger}(x)f(\bar{z}), & -\alpha<\tau<0.
\end{array}\right.
\end{equation}
The conditions on the function $f(\bar{z})=f(\tau-ix)$ are as follows\\
(i) $f$ is antianalytical in the complement of the two vertical 
(i.e. parallel to $x$ axis) branch cuts at $\tau=0$ and  $\tau=-\alpha$\\ 
(ii) $f(\infty)=1$\\
(iii)  Values of $f$ on the opposite sides of the cuts are related by
\begin{eqnarray}
\label{e-13}
f(-0-\alpha-ix) &=& S^{\dagger}(x)f(+0-\alpha-ix)\nonumber\\
S^{\dagger}(x)f(-0-ix) &=& f(+0-ix).
\end{eqnarray}
Conditions (i)-(iii) describe a matrix Riemann-Hilbert problem.
Once the decomposition~(\ref{e-9}) is found  the
solution of~(\ref{e-49}) is given by the Wess-Zumino functional
\begin{eqnarray}
\label{e-11}
\tilde{\mathcal W}(A) &=& -\frac{1}{8\pi}\int dx d\tau\tr\{\partial_{\mu}g g^{-1}
\partial_{\mu}g g^{-1}\}\nonumber\\
 &-& \frac{i}{12\pi}\int dx d\tau dy \epsilon^{\mu\nu\lambda}
\tr\{\partial_{\mu}g g^{-1} \partial_{\nu}g g^{-1}
\partial_{\lambda}g g^{-1}\}. 
\end{eqnarray}
In  the second integral   $g$ is continued on the membrane that has
$g(\tau,x)$  as a boundary. 
  We choose the continuation $g(\tau,x,y)$ with $0<y<1$ based on an
arbitrary interpolation $S(x,y)$ between  $S(x)=S(x,1)$ and $1=S(x,0)$
\begin{equation} 
\label{e-14}
g(\tau,x,y)=\left\{
\begin{array}{ll}
f(\bar{z},y), & \tau<-\alpha,\tau>0\\
S^{\dagger}(x,y)f(\bar{z},y), & -\alpha<\tau<0.
\end{array}\right.
\end{equation}
The function  $f(\bar{z},y)$  has the same branch cuts  and
behavior at infinity as $f(\bar{z})$ and obeys~(\ref{e-13})  
at each value of $y$
\begin{eqnarray}
\label{e-15}
f(-0-\alpha-ix,y) &=& S^{-1}(x,y)f(+0-\alpha-ix,y)\nonumber\\
S^{-1}(x,y)f(-0-ix,y) &=& f(+0-ix,y).
\end{eqnarray} 
Substituting~(\ref{e-14}) into ~(\ref{e-11}) and taking into account~(\ref{e-74}) we get 
(see appendix~\ref{s-64} for details) 
\begin{eqnarray}
\label{e-21}
\mathcal W &=& -\frac{i}{4\pi}\int_{\beta_{1}}\tr\{dS S^{-1}
df(-\alpha+0-ix,y)f^{-1}\}-
\frac{i}{4\pi}\int_{\beta_{2}}\tr\{dS S^{-1}
df(-0-ix,y)f^{-1}\}-\nonumber\\
&-& \frac{\alpha}{4\pi}
\int dx \tr\{\partial_{x}S(x) S^{-1}(x)
\partial_{x}S(x) S^{-1}(x)\}.
\end{eqnarray}
where $\beta_{1}$ is the plane $\tau=-\alpha$ and $\beta_{2}$ is the
plane  $\tau=0$. Note that the last term in~(\ref{e-21}) exactly
cancels the contribution from the first order diagram in~(\ref{e-54}).  

\begin{figure}[h]
\epsfbox{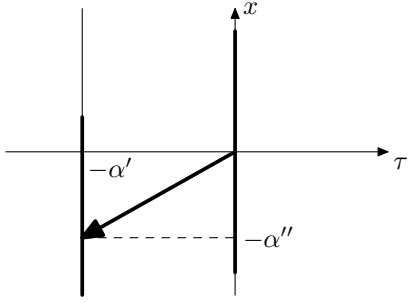}
\caption[]{\label{f-conti} Branch cuts for the continuation of
$f(\bar{z},y)$ to the complex values of $\alpha$. }
\end{figure}

\label{s-40}
The only remaining step  is
the analytic continuation of~(\ref{e-21})  from real positive $\alpha$
 to $\alpha=i\lambda$. The method of the continuation is illustrated on
 Fig.~\ref{f-conti}.
 We let $\alpha$  take the complex value  $\alpha=\alpha'+i\alpha''$
and obtain the continuation of~(\ref{e-14}) in the form 
\begin{equation}
\label{e-25}
g(\tau,x,y)=\left\{
\begin{array}{ll}
 f(\bar{z},y), & \tau<-\alpha',\tau>0 \\
 S^{-1}(x+\frac{\alpha''}{\alpha'}\tau,y)f(\bar{z},y), &
 -\alpha'<\tau<0
\end{array}
\right.
\end{equation}
where
\begin{eqnarray}
\label{e-24}
f(-0-\alpha'-ix,y) &=& S^{-1}(x-\alpha'',y)f(+0-\alpha'-ix)\nonumber\\
S^{-1}(x,y)f(-0-ix,y) &=& f(+0-ix,y).
\end{eqnarray}

  In the limiting case $\alpha=i\lambda$ the surfaces
$\tau=-\alpha'$ , $\tau=0$ coincide and~(\ref{e-21}) takes the
form~(\ref{eq:RH}).

\section{Energy dissipation in a quantum point contact}
\label{s-72}
In this section we apply Eq.~(\ref{eq:RH}) to a
biased  quantum point  contact.
The contact is described by the Hamiltonian 
\begin{equation}
\label{e-35}
H=-\frac{1}{2}\frac{\partial^{2}}{\partial x^{2}}+U(x),
\end{equation}
where the potential  $U(x)$ is
  local in space and therefore the scattering matrix  
\begin{equation}
S_{0}=\left[ 
 \begin{array}{cc}
  A & -B^{*}\\
  B & A^{*}
 \end{array}
\right]
\end{equation}
is almost energy independent.
The presence of the bias potential $V_b(t)$ is taken into account by  
adding the term $eV_b(t)\theta(x)$ to the Hamiltonian~(\ref{e-35}). 
The gauge
transformation  $\psi(x,t)\rightarrow\psi(x,t)e^{-i\phi(t)\theta(x)}$
where $\phi(t)=\int_{-\infty}^{t}eV(\tau)d\tau$
makes the Hamiltonian local in space
\begin{equation}
H(t)=\frac{1}{2}\left(-i\frac{\partial}{\partial x}-
\delta(x)\phi(t)\right)^2+U(x)
\end{equation}
with the scattering matrix given by~(\ref{eq:2x2S}).  

In what follows, we consider the rectangular bias pulse with amplitude $V_0$ and
duration $T_s$. The flux $\phi(t)$ is given by
\begin{equation}
\label{e-66}
\phi(t)=\left\{
 \begin{array}{ll}
  0, & t<0 \\	
  e V_0 t, & 0<t<T_s\\
  e V_0 T_s,& t>T_s.
 \end{array}
\right.
\end{equation} 	
The RH problem for the scattering matrix~(\ref{eq:2x2S})
can be solved in two limiting cases: when the contact is
transparent (i.e. does not reflect electrons) and when the time $T_s$ is
large.

\subsection{Transparent contact}
\label{s-trcont}
In the case of transparent contact the scattering
matrix~(\ref{eq:2x2S}) acquires the form 
\begin{equation}
\label{e-65}
S(t)=\left[\begin{array}{ll}
	  e^{i \phi(t)} & 0 \\
	  0 &  e^{-i\phi(t)}
     \end{array}\right].
\end{equation}
Since it remains diagonal at all times Eq.~(\ref{e-42}) can be used,
which gives
\begin{eqnarray}
\label{e-61}
\ln\chi(\lambda) &=& -\left(\frac{eV_0}{2\pi}\right)^2
\left[(\lambda-T_s)^2\ln\left(i\frac{\lambda}{T_s}-i\right)+\right. \nonumber\\ 
 &+& \left.(\lambda+T_s)^2\ln\left(i\frac{\lambda}{T_s}+i\right)
-2\lambda^2\ln\left(i\frac{\lambda}{T_s}\right)\right]
\end{eqnarray}

\begin{figure}[h]
\epsfbox{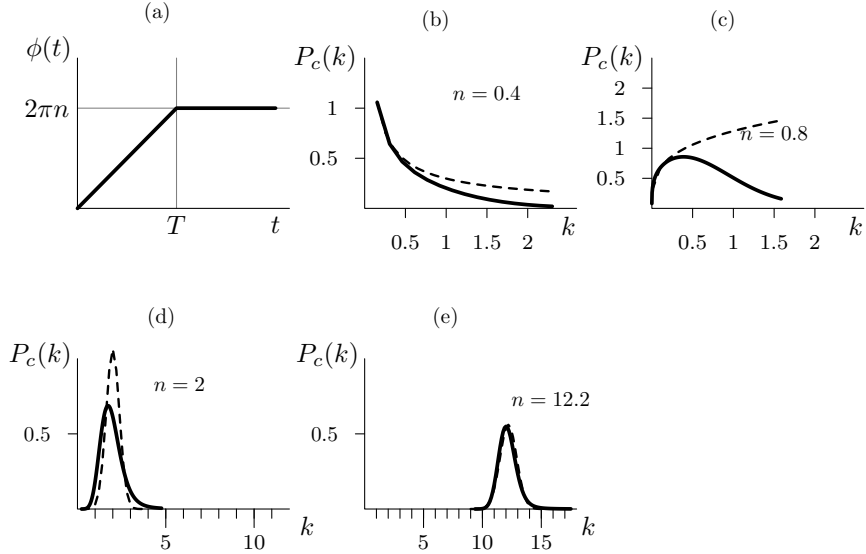}
\caption[]{\label{f-1} $P_c(k)$ is the probability that the one-dimensional
  Fermi gas in the wire acquires the energy $E=k eV_0$ after applying
  the voltage $V_0$ during the time $T_s$.
   The distribution $P_c(k)$
  is parametrised by a single parameter $n=e V_0 T_s/(2 \pi)$, which
  is the expected number of electrons passed through the contact. 
    \\[1ex]
  (a) The Faraday flux $\phi(t)$ as a function of time. The potential switches on
  at $t=0$  and switches off at $t=T_s$. \\
  (b) $P_c(k)$ for $n=0.4$ shows the Fermi-Edge singularity (the broken
  curve is the power law asymptote $P_c = C_{n}/k^{1-2n^2}$ with $C_n$ given
  by Eq.~(\ref{eq:ND-C}).\\
  (c) For $n=0.8>1/\sqrt{2}$ the behaviour at small $k$ is also determined by
  the power law asymptote (broken curve). However, now the charge distribution is
  not divergent at small $k$.\\      	
  (d), (e) $P_c(k)$ for the larger values of $n$ crosses over to the Gaussian
  distribution (shown by the broken curve) with the mean value $n$ and
  the standard deviation $\sigma = \frac{2}{\pi} \sqrt{\ln n}$.}
\end{figure}

Classically all  electrons in the energy strip $(E_F, E_F-e V_0)$
that can reach the contact are transmitted. The average number of such
electrons is  $n= \nu e V_0 v_F T_s= e V_0 T_s/(2\pi)$, where $v_F$ is
a Fermi velocity and $\nu=1/(2\pi v_F)$ is the density of states per
unit length. Since each transmitted electron contributes energy $e V_0$ to the
total dissipated energy
it is convenient to  introduce a  new rescaled  distribution function 
\begin{equation}
  \label{eq:P(k)}
  P_c(k) = e V_0 P(e V_0 k )
\end{equation}
that  measures energy in the units of $e V_0$. 
The new distribution $P_c(k)$ depends  only on the parameter
$n$. We plot $P_c(k)$
 for  three different values of $n$  in Fig~\ref{f-1}.  For small $n$
the distribution function has a divergence at small $k$
\begin{equation}
  \label{eq:n<1}
  P_c = C_n/k^{1-2 n^2} \hbox{ at } k \ll 1/n,
\end{equation}
which is the Nozieres and De Dominicis asymptote~(\ref{eq:ND-asymptote})
with the prefactor $C_n$ given by
\begin{equation}
  \label{eq:ND-C}
  C_n= 2 n \sin (2\pi n^2) e^{-3n^2} \Gamma(1-2 n^2) (2 \pi n)^{2 n^2-1}.
\end{equation}
In the other limit $n \gg 1$ the distribution looks like Gaussian with
mean value $n$ and the standard deviation $\sigma =\frac{2}{\pi}\sqrt{ \ln
n}$. The Gaussian approximation is valid only in the region $k-n\sim 1$.
The distribution function at $k \gg n$ is determined by the term
$\lambda^2\ln \lambda$ in~(\ref{e-61}), which leads to the logarithmically 
divergent second cumulant.

\subsection{Large $T_s$ asymptote of $\chi(\lambda)$}
\label{s-45}
For a semi-transparent contact we can
 still solve Eq.~(\ref{eq:fpm}) asymptotically when $T_s \gg 1/(e V_0)$. 
The  matrix $Q=S^{\dagger}(x+\lambda,y)S(x,y)$  has the form
\begin{equation}
Q=\left[
\begin{array}{ll}
  |A|^{2}e^{-i(\phi_{+}-\phi_{-})}+|B|^{2} & 
  -A^{*}B^{*}(e^{-i\phi_{+}}-e^{-i\phi_{-}})\\   
AB(e^{i\phi_{+}}-e^{i\phi_{-}}) &
|A|^{2}e^{i(\phi_{+}-\phi_{-})}+|B|^{2}  
\end{array}
\right],
\end{equation}
where $\phi_{+}=y\phi(x+\lambda)$,  $\phi_{-}=y\phi(x)$. 
In the limit $\lambda\ll T_s$ it has the simple time dependence  
\begin{equation}
Q=Q_{0}+Q_{+}e^{iyeV_0 x}+Q_{-}e^{-iyeV_0 x}
\end{equation}
where $Q_{0}$ is diagonal and $Q_{-}$($Q_{+}$) are upper (lower) triangular
 matrices, which are almost time-independent. This
 suggest the following anzats for $f_{\pm}$
\begin{eqnarray}
\label{e-69}
f_{+} &=& f_{0+}+f_{1+}e^{-yeV_0\bar{z}}\nonumber\\
f_{-} &=& f_{0-}+f_{1-}e^{yeV_0\bar{z}}.
\end{eqnarray}
After substituting~(\ref{e-69})  into~(\ref{eq:fpm}) we get
the approximate solution
\begin{eqnarray}      
f_{+} &=& (1+Q_{+}Q_{c}^{-1}e^{-yeV_0\bar{z}})f_{0+}\\
\label{eq:f-}
f_{-} &=& (1-Q_{0}^{-1}Q_{-}e^{yeV_0\bar{z}})f_{0-}
\end{eqnarray}
where $f_{0+}$,$f_{0-}$ are  diagonal matrices satisfying
the  equation
\begin{equation}
\label{eq:simplconj} 
f_{0+}f_{0-}^{-1}=Q_{c}
\end{equation}
with
\begin{equation} 
Q_{c}=Q_{0}-Q_{+}Q_{0}^{-1}Q_{-}=\left(
\begin{array}{cc}
|A|^{2}e^{-i(\phi_{+}-\phi_{-})}+|B|^{2} & 0\\
0 & \frac{1}{|A|^{2}e^{-i(\phi_{+}-\phi_{-})}+|B|^{2}}
\end{array}
\right).
\end{equation}
The solution for $\ln f_{0-}$ is given by
\begin{equation}
\ln f_{0-}=\frac{1}{2\pi i}\int 
\frac{\ln\left(|A|^{2}e^{-i(\phi_{+}(x')-\phi_{-}(x'))}+|B|^{2}\right)}{x'-x+i0}
dx' \left[\begin{array}{ll} 1 & 0\\ 0 & -1 \end{array}\right].
\end{equation}
Substituting $f_{-}=A_{-}f_{0-}$, where $A_{-}=1-Q_{0}^{-1}Q_{-}$, 
into~(\ref{eq:RH}) we get
\begin{eqnarray}
\mathcal{W} &=& \frac{i}{4\pi}\int\tr\left\{dS_{+}S_{+}^{-1}\wedge
dS_{-}S_{-}^{-1}\right\}
 - \frac{i}{4\pi}\int_{d}
\tr\left\{Q^{-1}dQ\wedge dA_{-}A_{-}^{-1}\right\}- \nonumber\\
&-& 
\frac{i}{4\pi}\int_{d}
\tr\left\{Q^{-1}dQ\wedge A_{-}(d\ln f_{0-})A_{-}^{-1}\right\}.
\end{eqnarray}
The first two integrals are proportional to $T_s$, the
last integral is proportional to $\ln(T_s/\lambda)$.
After long but straightforward algebra we finally
get~(\ref{eq:largeTans}).

\section{Conclusions}
We considered the energy dissipation in a Fermi gas subjected to a
perturbation with time dependent scattering matrix of a general form.
Result~(\ref{eq:RH}) relates the dissipated energy distribution
function $P(E)$ to a solution of the matrix Riemann-Hilbert
problem~(\ref{eq:fpm}).  

For a quantum point contact biased by a rectangular voltage pulse we
computed $P(E)$ analytically for long pulse durations and
analysed the relation between $P(E)$ and distribution of transmitted
charge. 
   We also obtained second and
third  cumulants  of the dissipated energy for a general
switching process. 

\section{Acknowledgments}

 We are grateful to  L.Levitov and D.Novikov for useful discussions. The work was
supported by the EPSRC grant GR/L91696.
\appendix
\section{One-particle scattering problem}
\label{s-1part}

In this section we relate the electron annihilation operator $\hat c_\kp$
in the Heisenberg representation to the one-particle scattering matrix
$S(E,t)$ and derive
Eqs.~(\ref{e-44},~\ref{e-43}) from the overview.

Solving the equation for the time evolution of
$\hat{c}_{\kp}$ 
\begin{equation}
\label{e-1}
 \dot{\hat{c}}_{\ve{k}}=-i\epsilon_{\ve{k}}\hat{c}_{\kp}-
i\sum_{\ve{k'}}V_{\kp\kd}(t)\hat{c}_{\kd}
\end{equation}  
perturbatively we get
\begin{eqnarray}
\label{e-71}
\hat{c}_{\kp}(t) &=& e^{-i\epsilon_\kp t}\left(c_\kp -
i\int_{-\infty}^{t}dt' V_{\kp\kd}(t')e^{i(\epsilon_\kp-\epsilon_\kd)t'}
c_\kd
+i^2\int_{-\infty}^{t}dt'V_{\kp\kd}(t')e^{i(\epsilon_\kp-\epsilon_\kd)t'} 
\times\right.\nonumber\\
&\times& \left. \int_{-\infty}^{t'}dt''
V_{\kd\ve{k''}}(t')e^{i(\epsilon_\kd-\epsilon_\ve{k''})t''} c_\ve{k''}- \dots \right). 
\end{eqnarray}
In the limit $t\to\infty$ we get
\begin{equation}
\label{e-46}
\hat{c}_\kp(t_\infty)=e^{-i\epsilon_{\kp}t_\infty}\sum_{\kd}(\delta_{\kp\kd}
-iF_{\kp\kd}(\epsilon_{\kp},\epsilon_{\kd}))c_{\kd}. 
\end{equation}
Here we introduced the function $F_{\kp\kd}(\epsilon,\epsilon')$
satisfying  the integral equation
\begin{equation}
\label{e-4}
F_{\kp\kd}(\epsilon,\epsilon') = V_{\kp\kd}(\epsilon-\epsilon')+
\sum_{\ve{k_1}}\int \frac{d\epsilon_1}{2\pi}
V_{\kp\ve{k_1}}(\epsilon-\epsilon_{1})G_{\ve{k_1}}(\epsilon_{1})
F_{\ve{k_1}\kd}(\epsilon_{1},\epsilon'),
\end{equation}
where $V_{\kp\kd}(\omega)=\int dt V_{\kp\kd}(t)e^{-i\omega t}$
and $G_{\kp}(\epsilon)=\frac{1}{\epsilon-\epsilon_{\kp}+i0}$ is the
retarded Green's function. To take into account the slowness of $V_{\kp\kd}(t)$ 
we introduce the new variables $E=(\epsilon+\epsilon')/2$ and
$\omega=\epsilon-\epsilon'$ and notice that $F$ varies slowly  when
$E$ changes but  varies fast when $\omega$ changes. Using Wigner's
representation
$F_{\kp\kd}(E,t)=\int e^{-i\omega t}F_{\kp\kd}(E,\omega)\frac{d\omega}{2\pi}$
we  get the  approximate solution of~(\ref{e-4}) 
\begin{eqnarray}
\label{e-5}
F_{\kp\kd}(E,t) & = & V_{\kp\kd}(t)+
\sum_{\ve{k_1}}V_{\kp\ve{k_1}}G_{\ve{k_1}}(E)F_{\ve{k_1}\kd}(E,t).
\end{eqnarray}
The first order correction to~(\ref{e-5}) is given by
\begin{equation}
\label{e-41}
 \delta F= -\frac{i}{2}\frac{\partial V}{\partial t}
\frac{\partial G}{\partial E}F+
\frac{i}{2}V\frac{\partial G}{\partial E}\frac{\partial F}{\partial t}
-\frac{i}{2}\frac{\partial V}{\partial t}G
\frac{\partial F}{\partial E}.
\end{equation}

   Eq.~(\ref{e-5}) is similar to the equation that one gets in
quantum scattering theory(see \cite{landau3}\S130). 
 
Substituting $F_{\kp\kd}$ given by~(\ref{e-5}) into~(\ref{e-46})
and neglecting the  terms of the order
 $\frac{\partial S}{\partial t}\frac{\partial S}{\partial E}$ we
get~(\ref{e-43}).

\section{Linked cluster expansion for $\chi(\lambda)$}
\label{s-68}
From Eq.~(\ref{eq:wkp}) we see that the characteristic
function~(\ref{e-chi-lambda})  is given by the linked cluster
expansion   
\begin{equation}
\label{e-diagrams}
\ln \chi = \ln \langle \Psi_0|e^{-i \hat{H}(\infty)\lambda}e^{i
  H_{0}\lambda}| \Psi_0 \rangle=-\myloop{10}{6}{12}-
\frac{1}{2}\myloop{4}{6}{12}-\frac{1}{3}\myloop{4}{4}{12}\cdots,
\end{equation}
where the solid line is the Green's function $G_{\kp}(\tau_2-\tau_1)$
and the vertex is $W_{\kp\kd}(\theta(\tau)-\theta(\tau-\lambda))$.
An integration with respect to fictious time $\tau$ is implied in each
vertex and summation with respect to $\kp$ in each line.

The causal Green's function is given by the sum
$$ G_\kp=G^R\theta(\epsilon_{\kp})+G^A\theta(-\epsilon_{\kp})$$
where $G^{R}(G^{A})$ are the retarded and advanced Green's function respectively
$$
G^{R}_\kp=-ie^{-i\epsilon_{\kp}(\tau_2-\tau_1)}
\theta(\tau_2-\tau_1)
$$
and
$$
G^{A}_\kp=ie^{-i\epsilon_{\kp}(\tau_2-\tau_1)}\theta(\tau_1-\tau_2).
$$

In the diagrams of second and higher orders  the 
main contribution in the sum over $\kp$ comes from the region near the
Fermi surface.  For example, in
the second order diagram we have
\begin{equation} 
\myloopII{}{} = \quad \myloopII{G^R}{G^R}
   +  \myloopII{G^R}{G^A}  + 
    \myloopII{G^A}{G^R}  +
   \myloopII{G^A}{G^A}
\end{equation}
since the causal  Green's function is a sum of advanced
and retarded ones.
 The
diagrams  with only advanced or only retarded
lines are equal to zero because of incompatible conditions set out by
the time $\theta$-functions. In the diagrams with both retarded and
advanced lines  because $W_{\kp\kd}$ is large only if 
$\epsilon_{\kp}-\epsilon_{\kd} < \frac{1}{T_s}$ all energies are close
to each other  and therefore close to the Fermi surface.
 This allows us to substitute
  $S(E_F,t)=S(t)$ for $S(E,t)$ in Eq.~(\ref{eq:wkp}). 
We notice that 
$$
\int \frac{d^{d}\kp}{(2\pi)^{d}} =
\frac{\Gamma(\frac{d}{2})\nu(E)}{\pi^{d/2-1}}\int
\frac{d\epsilon_{|\kp|}}{2\pi} \int d\ve{n}.
$$
  As the Green's functions do not depend on
direction of $\kp$  the integration over $\ve{n}=\kp/|\kp|$ involves only indexes
of scattering matrix in each vertex. This integration can be replaced
by summation over scattering channels if we  use an appropriate basis for
the scattering matrix.

  The first diagram in~(\ref{e-diagrams}) requires special attention
because the contributions from all the region under the Fermi surface
is significant and the energy dependence of $S(E,t)$ can not be neglected.
 We notice, however, that this diagram is proportional to the average
absorbed energy, because
\begin{equation}
\label{eq:first}
-\myloop{10}{6}{12}= -i\lambda \langle 0| \hat{H}_\infty - H_0 | 0 \rangle=
-i\lambda \langle E \rangle
\end{equation}
The average absorbed energy was computed in \cite{Makoshi:83} and is
given by~(\ref{eq:loss}).
Redefining the Green's function and the vertex in the diagram
expansion~(\ref{e-diagrams}) in a more convenient way we
get~(\ref{e-diagrams2}).  

\section{Integrating of continued $g$}
\label{s-64}
 The integral in~(\ref{e-11}) with $g$ given by~(\ref{e-14},\ref{e-15}) splits
into two parts: the surface integral 
\begin{equation}
\Omega(g)=-\frac{1}{8\pi}\int\tr(\partial_{\mu}gg^{-1}
\partial_{\mu}gg^{-1})d\tau d x
\end{equation}
and the volume integral
\begin{equation}
\label{e-56}
\Gamma(g)=-\frac{i}{12\pi}\int\epsilon^{\mu\nu\lambda}
\tr(\partial_{\mu}g g^{-1}\partial_{\nu}g g^{-1}
\partial_{\lambda}g g^{-1})d\tau d x d y.
\end{equation}

Firstly, we will deal with the surface integral. Because of the identity 
$\tr(\partial_{\mu}g g^{-1}\partial_{\mu}g g^{-1})=
4\tr(\partial_{z}g g^{-1}\partial_{\bar{z}}g g^{-1})$
the expression under the integral is zero in the region
where $\tau>0$ or $\tau<-\alpha$ where $\partial_z g=0$. Combining
\begin{eqnarray}
\tr(\partial_{\mu}(gh)h^{-1}g^{-1}\partial_{\mu}(gh)h^{-1}g^{-1}) &=& 
\tr(\partial_{\mu}g g^{-1}\partial_{\mu}g g^{-1})+\nonumber\\
+\tr(\partial_{\mu}h h^{-1}\partial_{\mu}h h^{-1}) &+&
2\tr(g^{-1}\partial_{\mu}g \partial_{\mu}h h^{-1})
\end{eqnarray}
 with Eq.~(\ref{e-12}) we get 
\begin{eqnarray}
\label{e-16}
\Omega(g) &=& -\frac{1}{8\pi}\int_{-\alpha}^{0} d\tau
\int dx \tr(\partial_{x}S^{-1}(x)S(x)
\partial_{x}S^{-1}(x)S(x))-\nonumber\\
 &-&\frac{1}{4\pi}\int_{-\alpha}^{0}d\tau
\int dx \tr(S \partial_{x}S^{-1}\partial_{x}f(\bar{z})
f^{-1}(\bar{z})).
\end{eqnarray}
Notice, that  
$\partial_{x}f=-i \partial_{\bar{z}}f$ due to antianalyticity of $f$.

In Eq.~(\ref{e-56}) the interior of the sphere is separated into
three parts by surfaces $\tau=-\alpha$ and $\tau=0$. 
Everywhere except in the middle part where $0<\tau<-\alpha$ 
the  differential 
form $\tr(dg g^{-1}\wedge dg g^{-1}\wedge dg g^{-1})$  is  zero
because in those regions $g$ depends 
 on two variables only. In the remaining part we have  
 $g=S^{-1}(x,y)f(\bar{z},y)$ and the differential form is exact
\begin{eqnarray}
\tr\{(dg g^{-1})^{3}\} &=& -\tr\{(S^{-1} dS)^{3}\}+
\tr\{(df f^{-1})^{3}\}\nonumber\\
 &+& 3d\tr(dS S^{-1}\wedge df f^{-1})=
3d\tr(dS S^{-1}\wedge df f^{-1}).
\end{eqnarray}
Applying  Stokes' theorem we get
\begin{eqnarray}
\label{e-57}
\Gamma(g) &=& -\frac{i}{4\pi}\int_{-\alpha}^{0}d\tau\int dx
\tr(S \partial_{x}S^{-1}\partial_{\bar{z}}f(\bar{z})f^{-1})-
\nonumber\\
 &-& \frac{i}{4\pi}\int_{\beta_{1}}\tr\{dS S^{-1}
df(-\alpha+0-ix,y)f^{-1}\}-\nonumber\\
 &-& \frac{i}{4\pi}\int_{\beta_{2}}\tr\{dS S^{-1}
df(-0-ix,y)f^{-1}\}
\end{eqnarray}
where $\beta_{1}$ is a surface $\tau=-\alpha$ and $\beta_{2}$ is a
surface $\tau=0$. The first term in~(\ref{e-57}) cancels the second one 
from~(\ref{e-16}) and thus we obtain~(\ref{e-21}).

\end{fmffile}


\begin{thebibliography}{10}
\bibitem{coords}
The coordinate representation is introduced by the formula
  $f(\tau,x)=\int\frac{d\xi}{2\pi}\frac{dp}{2\pi}e^{ipx}e^{i\xi\tau}f(\xi,p)$.
\end{thebibliography}
\end{document}